\documentclass[prd,onecolumn,nofootinbib,showpacs]{revtex4}
\usepackage{graphicx}
\usepackage{latexsym}
\usepackage{color}
\begin{document}

\title{Cosmological CPT Violation and CMB Polarization Measurements}

\author{Jun-Qing Xia}

\affiliation{Scuola Internazionale Superiore di Studi Avanzati, Via
Bonomea 265, I-34136 Trieste, Italy}

\date{\today}

\begin{abstract}

In this paper we study the possibility of testing Charge-Parity-Time
Reversal (CPT) symmetry with cosmic microwave background (CMB)
experiments. We consider two kinds of Chern-Simons (CS) term,
electromagnetic CS term and gravitational CS term, and study their
effects on the CMB polarization power spectra in detail. By
combining current CMB polarization measurements, the seven-year
WMAP, BOOMERanG 2003 and BICEP observations, we obtain a tight
constraint on the rotation angle $\Delta\alpha=-2.28\pm1.02$ deg
($1\,\sigma$), indicating a $2.2\,\sigma$ detection of the CPT
violation. Here, we particularly take the systematic errors of CMB
measurements into account. After adding the QUaD polarization data,
the constraint becomes $-1.34<\Delta\alpha<0.82$ deg at 95\%
confidence level. When comparing with the effect of electromagnetic
CS term, the gravitational CS term could only generate TB and EB
power spectra with much smaller amplitude. Therefore, the induced
parameter $\epsilon$ can not be constrained from the current
polarization data. Furthermore, we study the capabilities of future
CMB measurements, Planck and CMBPol, on the constraints of
$\Delta\alpha$ and $\epsilon$. We find that the constraint of
$\Delta\alpha$ can be significantly improved by a factor of $15$.
Therefore, if this rotation angle effect can not be taken into
account properly, the constraints of cosmological parameters will be
biased obviously. For the gravitational CS term, the future Planck
data still can not constrain $\epsilon$ very well, if the primordial
tensor perturbations are small, $r <0.1$. We need the more accurate
CMBPol experiment to give better constraint on $\epsilon$.

\end{abstract}

\pacs{98.80.Cq, 98.80.Es, 11.30.Cp, 11.30.Er}

\maketitle


\section{Introduction}

In the standard model of particle physics, Charge-Parity-Time
Reversal (CPT) symmetry is a fundamental symmetry. Probing its
violation is an important way to search for the new physics beyond
the standard model. Up to now, CPT symmetry has passed a number of
high-precision experimental tests and no definite signal of its
violation has been observed in the laboratory. So, the present CPT
violating effects, if they exist, should be very small to be
amenable to the experimental limits.

However, the CPT symmetry could be dynamically violated in the
expanding universe. In the literature \cite{Li:2002,Li:2003,
Li:2004,Feng:2005,Li:2007}, the cosmological CPT violation mechanism
has an interesting feature, which is that the CPT violating effects
at present are too small to be detected by the laboratory
experiments, but large enough in the early universe to account for
the generation of matter-antimatter asymmetry. More importantly,
these types of CPT violating effects could be accumulated to be
observable in the cosmological experiments
\cite{Feng:2005,Li:2007,Feng:2006}. With the accumulation of
high-quality observational data, especially those from the cosmic
microwave background (CMB) experiments, cosmological observation
becomes a powerful way to test the CPT symmetry.

In this paper we firstly consider the cosmological CPT violation in
the photon sector where the electrodynamics is modified by a
Chern-Simons term $\mathcal{L}_{\rm ecs}\sim
p_{\mu}A_{\nu}\tilde{F}^{\mu\nu}$. Here, $p_{\mu}$ is an external
vector and
$\tilde{F}^{\mu\nu}=(1/2)\epsilon^{\mu\nu\rho\sigma}F_{\rho\sigma}$
is the dual of the electromagnetic tensor. This term violates the
Lorentz and CPT symmetries if $p_{\mu}$ is treated as an external
field. One of the physical consequences of this electromagnetic
Chern-Simons (eCS) term is the rotation of the polarization
direction of electromagnetic waves propagating over large distances
\cite{Carroll:1990}, which is known as ``cosmological
birefringence''. This rotation angle $\Delta\alpha$ can be obtained
by observing polarized radiation from distant sources such as radio
galaxies, quasars, and CMB. For the standard theory of CMB, the TB
and EB cross-correlation power spectra vanish. In the presence of
the non-zero rotation angle $\Delta\alpha$, one expects to observe
non-zero TB and EB power spectra, even if they are zero at the last
scattering surface. Denoting the rotated quantities with a prime,
one gets \cite{Feng:2006,Lue:1999}
\begin{eqnarray}
C_{\ell}^{\rm 'TB} &=& C_{\ell}^{\rm TE}\sin(2\Delta\alpha)~, \nonumber\\
C_{\ell}^{\rm 'EB} &=&
\frac{1}{2}(C_{\ell}^{\rm EE}-C_{\ell}^{\rm BB})\sin(4\Delta\alpha)~,\nonumber\\
C_{\ell}^{\rm 'TE} &=& C_{\ell}^{\rm TE}\cos(2\Delta\alpha)~,\nonumber\\
C_{\ell}^{\rm 'EE} &=& C_{\ell}^{\rm EE}\cos^2(2\Delta\alpha) +
C_{\ell}^{\rm BB}\sin^2(2\Delta\alpha)~,\nonumber\\
C_{\ell}^{\rm 'BB} &=& C_{\ell}^{\rm BB}\cos^2(2\Delta\alpha) +
C_{\ell}^{\rm EE}\sin^2(2\Delta\alpha)~,
\end{eqnarray}
while the CMB temperature power spectrum remains unchanged. Hence we
can use the CMB polarization measurements to test the Lorentz and
CPT symmetries in this eCS model (see
Refs.\cite{Lue:1999,Feng:2005,Feng:2006,Li:2007,Xia:2008a,WMAP5,
Xia:2008b,Wu:2009,Brown:2009,WMAP7,Xia:2010,Liu:2006,Xia:Planck,Others},
and references within). With homogeneous and isotropic rotation
angle, Ref.\cite{Xia:2010} used the newly released CMB observations
and found that a non-zero rotation angle $\Delta\alpha=-2.2\pm0.8$
deg ($1\,\sigma$) is favored by the CMB polarization data from the
seven-year WMAP (WMAP7) \cite{WMAP7}, BOOMERanG 2003 (B03)
\cite{B03} and BICEP \cite{Bicep} observations.

Besides the eCS term, the gravitational Chern-Simons (gCS) term
could also generate the non-zero CMB TB and EB cross-correlation
power spectra. Following
Refs.\cite{Saito:2007,Li:2009,Gluscevic:2010}, here we consider a
Lorentz and CPT violating term in the gravity sector
$\mathcal{L}_{\rm gcs}\sim
\epsilon^{\mu\nu\rho\sigma}R^\alpha_{\beta\mu\nu}R^\beta_{\alpha\rho\sigma}$,
where $R^\alpha_{\beta\mu\nu}$ is the Riemann tensor. This gCS term
does not affect the evolution of background and scalar
perturbations. Therefore, the effect of gCS term only appears in the
evolution of tensor perturbations, if we neglect the vector
perturbations. As we know, the gravitational wave has two
independent polarized components denoted by $+$ and $\times$. We
usually use the right- and left-handed circular polarized
components:
\begin{equation}
h^{\rm R}=\frac{1}{\sqrt{2}}(h^{+}-ih^{\times})~,~~~h^{\rm
L}=\frac{1}{\sqrt{2}}(h^{+}+ih^{\times})~.
\end{equation}
The power spectra for different handedness are defined as:
$\langle{h^{\rm s*}({\bf k_1})h^{\rm s'}({\bf k_2})}\rangle=P^{\rm
ss'}_{h}\delta^3({\bf k_1}-{\bf k_2})\delta_{\rm ss'}$, where
superscripts $s$ and $s'$ stand for the circular polarized state,
$s,\,s'={\rm R,\,L}$. Due to the presence of this gCS term, the
produced power spectra $P^{\rm R}_h$ and $P^{\rm L}_h$ are not
equal. Similarly with Refs.\cite{Saito:2007,Li:2009,Gluscevic:2010},
we use a parameter $\epsilon$ to characterize this discrepancy:
\begin{eqnarray}
P^{\rm R}_h=\frac{1}{2}P_h(1-\epsilon)~&,&~~~P^{\rm
L}_h=\frac{1}{2}P_h(1+\epsilon)~,\nonumber\\
P^{\rm R}_h+P^{\rm L}_h=P_h~&,&~~~P^{\rm R}_h-P^{\rm L}_h=-\epsilon
P_h~.
\end{eqnarray}
We can see that $\epsilon=-1,\,0,\,1$ denote purely right-handed
polarized, unpolarized and purely left-handed polarized
gravitational wave, respectively. In this paper we only consider the
scale-independent $\epsilon$ simply.

The CMB power spectra generated by the tensor perturbations are
given by:
\begin{eqnarray}
C_\ell^{\rm XX'}&=&(4\pi)^2\int k^2dk[P^{\rm R}_h(k)+P^{\rm
L}_h(k)]\Delta^{\rm X}_\ell(k)\Delta^{\rm X'}_\ell(k)=(4\pi)^2\int
k^2dkP_h(k)\Delta^{\rm X}_\ell(k)\Delta^{\rm X'}_\ell(k)~,\nonumber\\
C_\ell^{\rm YY'}&=&(4\pi)^2\int k^2dk[P^{\rm R}_h(k)-P^{\rm
L}_h(k)]\Delta^{\rm Y}_\ell(k)\Delta^{\rm
Y'}_\ell(k)=-(4\pi)^2\epsilon\int k^2dkP_h(k)\Delta^{\rm
Y}_\ell(k)\Delta^{\rm Y'}_\ell(k)~,
\end{eqnarray}
where ${\rm XX'}$ and ${\rm YY'}$ denote TT, TE, EE, BB and TB, EB,
respectively, and $\Delta^{\rm X}_\ell(k)$ are the transfer
functions. These equations show that the TT, TE, EE and BB power
spectra only depend on the sum of the primordial power spectra of
gravitational waves, while the TB and EB power spectra rely on the
difference between the power spectra of right- and left-handed
polarized components $-\epsilon P_h$. Therefore, in the presence of
the gCS term, the non-zero TB and EB correlations will be generated
and other four power spectra are unchanged.

In this paper we will study the effects of eCS and gCS terms on the
CMB power spectra in detail and perform a global analysis on them by
using the latest CMB polarization measurements, as well as the
future simulated CMB data. The structure of the paper is as follows:
in section \ref{data} we describe the current and future simulated
datasets we use. Section \ref{result} contains our main results from
the current observations and future measurements, while section
\ref{summary} is dedicated to the conclusions and discussion.


\section{CMB Datasets}\label{data}

\subsection{Current Datasets}

In our calculations we mainly use the full data of WMAP7 temperature
and polarization power spectra \cite{WMAP7}. The WMAP7 polarization
data are composed of TE/TB/EE/BB/EB power spectra on large scales
($2\leq \ell\leq23$) and TE/TB power spectra on small scales
($24\leq \ell \leq800$), while the WMAP7 temperature data are only
used to set the underlying cosmology. For the systematic error, the
WMAP instrument can measure the polarization angle to within
$\pm1.5$ deg of the design orientation \cite{WMAP7Page}. In the
computation we use the routines for computing the likelihood
supplied by the WMAP team. Besides the WMAP7 information, we also
use some small-scale CMB observations.

The {\it BOOMERanG dated January 2003 Antarctic flight} \cite{B03}
measures the small-scale CMB polarization power spectra in the range
of $150\leq \ell\leq1000$. Recently, the BOOMERanG collaboration
re-analyzed the CMB power spectra and took into account the effect
of systematic errors rotating the polarization angle by $-0.9\pm0.7$
deg \cite{Pagano:2009}.

Recently, the {\it Background Imaging of Cosmic Extragalactic
Polarization} (BICEP) \cite{Bicep} and {\it QU Extragalactic Survey
Telescope at DASI} (QUaD) \cite{Quad} collaborations released their
high precision data of the CMB temperature and polarization
including the TB and EB power spectra. These two experiments,
locating at the South Pole, are the bolometric polarimeters designed
to capture the CMB information at two different frequency bands of
$100$GHz and $150$GHz, and on small scales -- the released first
two-year BICEP data are in the range of $21\leq \ell\leq335$
\cite{Bicep}; whereas the QUaD team measures the polarization
spectra at $164\leq \ell\leq 2026$, based on an analysis of the
observation in the second and third season
\cite{Wu:2009,Brown:2009}. They also provide the systematic errors
of measuring the polarization angle, $\pm0.7$ deg and $\pm0.5$ deg,
for BICEP and QUaD observations, respectively.

\begin{table}[t]
\caption{Assumed experimental specifications. We use the CMB power
spectra only at $l\leq2500$. The noise parameters $\Delta_T$ and
$\Delta_P$ are given in units of $\mu$K-arcmin.}\label{futureCMB}
\begin{center}
\begin{tabular}{lcccccc}
\hline \hline

~Experiment~ & ~$f_{\rm sky}$~ & ~$\ell_{\rm max}$~ & ~(GHz)~ &
~$\theta_{\rm FWHM}$~ & ~$\Delta_T$~ & ~$\Delta_P$~ \\

\hline

~PLANCK & 0.65 & 2500 & 100 & 9.5' &  6.8 & 10.9 \\
        &      &      & 143 & 7.1' &  6.0 & 11.4 \\
        &      &      & 217 & 5.0' & 13.1 & 26.7 \\
~CMBPol & 0.65 & 2500 & 217 & 3.0' &  1.0 &  1.4 \\

\hline \hline
\end{tabular}
\end{center}
\end{table}

\subsection{Future Datasets}

Current CMB measurements are still not accurate enough to verify the
possible CPT violation. We follow the method given in
Refs.\cite{Xia:2008a,Xia:Planck} and simulate the CMB power spectra
with the assumed experimental specifications of the future Planck
\cite{Planck} and CMBPol \cite{CMBPol} measurements. We choose the
best-fit model from the WMAP7 data \cite{WMAP7} as the fiducial
model.

In Table \ref{futureCMB} we list the assumed experimental
specifications of the future Planck and CMBPol measurements. The
likelihood function is $\mathcal{L}\propto \exp(-\chi_{\rm
eff}^2/2)$ and
\begin{equation}\label{simu}
\chi^2_{\rm eff}=\sum_{\ell}(2\ell+1)f_{\rm
sky}\left(\frac{A}{|\bar{C}|}+\ln\frac{|\bar{C}|}{|\hat{C}|}+3\right)~,
\end{equation}
where $f_{\rm sky}$ denotes the observed fraction of the sky in the
real experiments, $A$ is defined as:
\begin{eqnarray}\label{AAA}
A &=&
\hat{C}^{TT}_{\ell}(\bar{C}^{EE}_{\ell}\bar{C}^{BB}_{\ell}-(\bar{C}^{EB}_{\ell})^2)+\hat{C}^{TE}_{\ell}(\bar{C}^{TB}_{\ell}\bar{C}^{EB}_{\ell}-\bar{C}^{TE}_{\ell}\bar{C}^{BB}_{\ell})\nonumber\\
  &+& \hat{C}^{TB}_{\ell}(\bar{C}^{TE}_{\ell}\bar{C}^{EB}_{\ell}-\bar{C}^{TB}_{\ell}\bar{C}^{EE}_{\ell})+\hat{C}^{TE}_{\ell}(\bar{C}^{TB}_{\ell}\bar{C}^{EB}_{\ell}-\bar{C}^{TE}_{\ell}\bar{C}^{BB}_{\ell})\nonumber\\
  &+& \hat{C}^{EE}_{\ell}(\bar{C}^{TT}_{\ell}\bar{C}^{BB}_{\ell}-(\bar{C}^{TB}_{\ell})^2)+\hat{C}^{EB}_{\ell}(\bar{C}^{TE}_{\ell}\bar{C}^{TB}_{\ell}-\bar{C}^{TT}_{\ell}\bar{C}^{EB}_{\ell}) \nonumber\\
  &+& \hat{C}^{TB}_{\ell}(\bar{C}^{TE}_{\ell}\bar{C}^{EB}_{\ell}-\bar{C}^{EE}_{\ell}\bar{C}^{TB}_{\ell})+\hat{C}^{EB}_{\ell}(\bar{C}^{TE}_{\ell}\bar{C}^{TB}_{\ell}-\bar{C}^{TT}_{\ell}\bar{C}^{EB}_{\ell})\nonumber\\
  &+& \hat{C}^{BB}_{\ell}(\bar{C}^{TT}_{\ell}\bar{C}^{EE}_{\ell}-(\bar{C}^{TE}_{\ell})^2)~,
\end{eqnarray}
and $|\bar{C}|$ and $|\hat{C}|$ denote the determinants of the
theoretical and observed data covariance matrices respectively,
\begin{eqnarray}\label{CCC}
|\bar{C}|&=&\bar{C}^{TT}_{\ell}\bar{C}^{EE}_{\ell}\bar{C}^{BB}_{\ell}+2\bar{C}^{TE}_{\ell}\bar{C}^{TB}_{\ell}\bar{C}^{EB}_{\ell}
           -\bar{C}^{TT}_{\ell}(\bar{C}^{EB}_{\ell})^2\nonumber\\
         & &-\bar{C}^{EE}_{\ell}(\bar{C}^{TB}_{\ell})^2-\bar{C}^{BB}_{\ell}(\bar{C}^{TE}_{\ell})^2~,\nonumber\\
|\hat{C}|&=&\hat{C}^{TT}_{\ell}\hat{C}^{EE}_{\ell}\hat{C}^{BB}_{\ell}+2\hat{C}^{TE}_{\ell}\hat{C}^{TB}_{\ell}\hat{C}^{EB}_{\ell}
           -\hat{C}^{TT}_{\ell}(\hat{C}^{EB}_{\ell})^2\nonumber\\
         & &-\hat{C}^{EE}_{\ell}(\hat{C}^{TB}_{\ell})^2-\hat{C}^{BB}_{\ell}(\hat{C}^{TE}_{\ell})^2~.
\end{eqnarray}
The likelihood has been normalized with respect to the maximum
likelihood $\chi^2_{\rm eff}=0$, where $\bar{C}^{\rm
XY}_{\ell}=\hat{C}^{\rm XY}_{\ell}$.


\section{Numerical Results}\label{result}

In our study we make a global analysis to the CMB data with the
public available Markov Chain Monte Carlo package {\tt CosmoMC}
\cite{Lewis:2002}, which has been modified to compute the non-zero
TB and EB power spectra discussed above. We assume the purely
adiabatic initial conditions and impose the flatness condition
motivated by inflation. Our basic parameter space is: ${\bf P}
\equiv (\omega_{b}, \omega_{c}, \Omega_\Lambda, \tau, n_{s}, A_{s},
r)$, where $\omega_{b}\equiv\Omega_{b}h^{2}$ and
$\omega_{c}\equiv\Omega_{c}h^{2}$ are the physical baryon and cold
dark matter densities relative to the critical density,
$\Omega_\Lambda$ is the dark energy density relative to the critical
density, $\tau$ is the optical depth to re-ionization, $A_{s}$ and
$n_{s}$ characterize the primordial scalar power spectrum, $r$ is
the tensor to scalar ratio of the primordial spectrum. For the pivot
of the primordial spectrum we set $k_{\rm s0}=0.002\,$Mpc$^{-1}$.
Furthermore, in our analysis we include the CMB lensing effect,
which also produces B modes from E modes \cite{lensing}, when we
calculate the theoretical CMB power spectra.

\begin{table}[t]
\caption{Constraints on the rotation angle $\Delta\alpha$ (68\%
C.L.) from various CMB data combinations.}\label{table:eCScur}
\begin{tabular}{c|c|c}
\hline \hline
CMB Data sets & without systematics & with systematics \\
\hline

WMAP7&$-1.06\pm1.39$ deg&$-1.14\pm2.05$ deg\\
B03&$-5.97\pm4.05$ deg&$-4.63\pm4.16$ deg\\
BICEP&$-2.59\pm0.99$ deg&$-2.57\pm1.21$ deg\\
QUaD&$0.60\pm0.40$ deg&$0.59\pm0.64$ deg\\

\hline

WMAP7+B03+BICEP&$-2.23\pm0.83$ deg&$-2.28\pm1.02$ deg\\
WMAP7+BICEP+QUaD&$0.07\pm0.36$ deg&$-0.22\pm0.54$ deg\\
WMAP7+B03+BICEP+QUaD&$0.04\pm0.34$ deg&$-0.26\pm0.54$ deg\\

\hline  \hline
\end{tabular}
\end{table}

\subsection{Electromagnetic Chern-Simons Term}

Firstly, we consider the constraint on the rotation angle
$\Delta\alpha$, induced by the eCS term, from the current CMB
measurements. As we know, this rotation angle is accumulated along
the journey of CMB photons, and the constraints on the rotation
angle depends on the multipoles $\ell$ \cite{Liu:2006}.
Refs.\cite{WMAP5,WMAP7} found that the rotation angle is mainly
constrained from the high-$\ell$ polarization data, and the
polarization data at low multipoles do not affect the result
significantly. Therefore, in our analysis, we assume a constant
rotation angle $\Delta\alpha$ at all multipoles. Further, we also
impose a conservative flat prior on $\Delta\alpha$ as,
$-\pi/2\leq\Delta\alpha\leq\pi/2$.

In our previous works, we only presented the statistical errors of
rotation angle and did not consider the possible systematic errors
of CMB measurements. Inspired by Ref.\cite{Pagano:2009}, in this
paper we consider two rotation angles, $\Delta\alpha$ and $\beta$,
in order to take into account the real rotation signal and a
systematic error for each CMB polarization measurement. Therefore,
in our analyses we have five free parameters in this eCS model: the
rotation angle signal $\Delta\alpha$ and four systematic errors,
$\beta_{\rm WMAP7},\beta_{\rm B03},\beta_{\rm BICEP},\beta_{\rm
QUaD}$, for four CMB observations, respectively. And we impose
priors on these four systematic errors:
\begin{eqnarray}
\beta_{\rm WMAP7}=0.0 \pm 1.5~{\rm deg}&,&~~\beta_{\rm
B03}=-0.9 \pm 0.7~{\rm deg}~,\nonumber\\
\beta_{\rm BICEP}=0.0 \pm 0.7~{\rm deg}&,&~~\beta_{\rm QUaD}=0.0 \pm
0.5~{\rm deg}~,
\end{eqnarray}
and marginalize over them to constrain the rotation angle. In Table
\ref{table:eCScur} we present current constraints on $\Delta\alpha$
from the WMAP7, B03, BICEP and QUaD CMB polarization power spectra
with and without CMB systematic errors.

When only using WMAP7 power spectra data at all multipoles $\ell$,
without the systematic effect ($\beta_{\rm WMAP7}=0$), we obtain the
constraint on the rotation angle: $\Delta\alpha=-1.06\pm1.39$ deg at
$68\%$ confidence level, which is quite consistent with that
obtained from the WMAP team \cite{WMAP7} and is a significant
improvement over the WMAP3 \cite{Xia:2008a} and WMAP5
\cite{WMAP5,Xia:2008b} results. After adding the prior of
$\beta_{\rm WMAP7}$ into the $\chi^2$ calculation, the constraint
becomes weaker: $\Delta\alpha=-1.14\pm2.05$ deg ($68\%$ C.L.) and
$-5.24<\Delta\alpha<2.96$ deg ($95\%$ C.L.). Similarly, we revisit
the constraint on $\Delta\alpha$ from the B03 polarization data and
obtain the constraints at $68\%$ C.L.: $\Delta\alpha=-5.97\pm4.05$
deg and $\Delta\alpha=-4.63\pm4.16$ deg, without and with the CMB
systematic effect, respectively. The results are in good agreement
with the previous results of Refs.\cite{Feng:2006,Pagano:2009}.
These two CMB measurements show a weak indication for a non-zero
rotation angle about $1\,\sigma$ confidence level.

Recently, BICEP and QUaD collaborations also released their high
precision CMB polarization data. In our previous work
\cite{Xia:2010}, we found that the BICEP data alone could give very
tight constraint on the rotation angle: $\Delta\alpha=-2.59\pm0.99$
deg ($68\%$ C.L.), when fixing $\beta_{\rm BICEP}=0$, which means
that BICEP data alone favors a non-zero rotation angle at about
$2.5\,\sigma$ confidence level. When we include the systematic
effect of BICEP measurement, the constraints at $68\%$ and $95\%$
C.L. are:
\begin{eqnarray}
\Delta\alpha=-2.57\pm1.21~{\rm deg}~(68\%~{\rm C.L.})~,~~ -4.99 <
\Delta\alpha < -0.15~{\rm deg}~(95\%~{\rm C.L.})~,
\end{eqnarray}
which still gives a more than $2\,\sigma$ detection of a
non-vanishing rotation angle. Furthermore, when WMAP7 and B03 data
are added to the BICEP sample, the constraint on $\Delta\alpha$ gets
tightened:
\begin{eqnarray}
\Delta\alpha=-2.28\pm1.02~{\rm deg}~(68\%~{\rm C.L.})~,~~ -4.32 <
\Delta\alpha < -0.24~{\rm deg}~(95\%~{\rm C.L.})~,\label{threeconst}
\end{eqnarray}
which implies $\Delta\alpha\neq0$ at $2.2\,\sigma$ confidence level,
even considering systematic effects of these three CMB measurements.

Finally, we constrain the rotation angle from the QUaD polarization
data. Similarly with previous results of QUaD collaboration, we use
the QUaD data alone and obtain the constraint at 68\% confidence
level without and with systematic effect: $\Delta\alpha=0.60\pm0.40$
deg and $\Delta\alpha=0.59\pm0.64$ deg, respectively. When comparing
with the result of WMAP7+B03+BICEP [Eq.(\ref{threeconst})], there is
a $\sim2\,\sigma$ tension between QUaD and WMAP7+B03+BICEP
observations, no matter whether we include the systematic effects of
these CMB measurements. As we discussed before \cite{Xia:2010}, this
tension should be taken care of in the further investigation. By
combining WMAP7, BICEP and QUaD data and including their systematic
effects, we obtain the constraint: $\Delta\alpha=-0.22\pm0.54$ deg
($68\%$ C.L.) and $-1.30<\Delta\alpha<0.86$ deg ($95\%$ C.L.), which
is consistent with the result of Ref.\cite{WMAP7}. When adding the
B03 data into the analysis, the constraint on the rotation angle
does not change significantly, due to large error bars of B03
polarization data.

\begin{figure}[t]
\begin{center}
\includegraphics[scale=0.65]{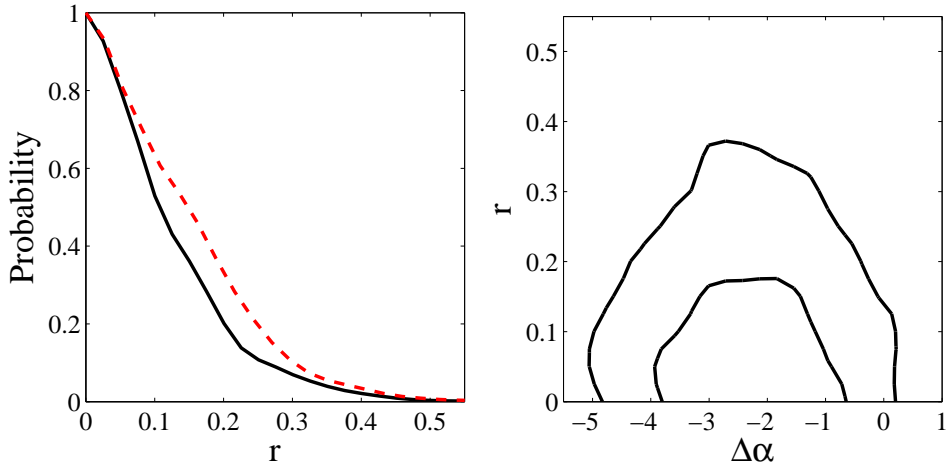}
\caption{Left panel: The one-dimensional posterior distributions of
the tensor to scalar ratio $r$ derived from the WMAP7+B03+BICEP data
combinations including (black solid line) and not including (red
dashed line) the rotation angle. Right panel: Two-dimensional
$1,2\,\sigma$ contours in the $\Delta\alpha-r$ plane.
\label{fig:tensor}}
\end{center}
\end{figure}

Moreover, the sources of the CMB polarization, especially for the
B-mode, are not unique. For example, the B-mode can be generated by
the cosmological birefringence as mentioned above; it might be
converted from E-mode by cosmic shear \cite{lensing}; it could be
the signature of the gravitational waves; or it can even be produced
by the instrumental systematics. Therefore, one should bear in mind
that the rotation angle might be degenerate with other cosmological
parameters or nuisance parameters when fitted to the polarization
data. Therefore, in order to distinguish these effects and obtain
the clean information of the primordial tensor B-mode, the rotation
angle has to be constrained, and the measurements of TB and EB power
spectra are really necessary. Otherwise, the constraints of
cosmological parameters might be biased, if this cosmological
birefringence effect can not be taken into account properly. In the
left panel of Fig.\ref{fig:tensor} we show the one-dimensional
posterior distributions of the tensor to scalar ratio $r$ derived
from the WMAP7+B03+BICEP data combinations. When including the
rotation angle, the 95\% upper limit of $r$ is: $r < 0.26$, while we
obtain $r < 0.29$ with $\Delta\alpha=0$. Since the data combination
WMAP7+B03+BICEP favors a negative rotation angle which contributes
the BB power spectrum, the BB power spectrum from primordial tensor
perturbations, as well as the upper limit of $r$, will be slightly
suppressed, consequently. We also show the two-dimensional contour
between the tensor and the rotation angle in the right panel of
Fig.\ref{fig:tensor}. Due to the large error bars of current CMB
polarization data, the effect of non-zero rotation angle is
neglectable on the constraints of cosmological parameters. However,
this effect might be important in the future forecasts.

\begin{table}[htbp]\footnotesize{
\caption{Constraints on the cosmological parameters (68\% C.L.) from
future CMB mock data.}\label{table:eCSfut}
\begin{tabular}{cccccccccc}
\hline \hline
Parameters&$100\,\Omega_bh^2$&$10\,\Omega_ch^2$&$10\,\Omega_\Lambda$&$10\,\tau$&$10\,n_s$&$\log[10^{10}A_s]$&$r(95\%~{\rm C.L.})$&$\Delta\alpha\,({\rm deg})$&$\Delta\chi^2$\\
Fiducial&$2.300$&$1.080$&$7.520$&$0.870$&$9.800$&$3.135$&${\rm See~below}$&${\rm See~below}$&$0$\\
\hline

\multicolumn{9}{c}{PLANCK, $r=0$, $\Delta\alpha$ free}\\
$\Delta\alpha=0^\circ$&$2.300\pm0.017$&$1.079\pm0.015$&$7.519\pm0.077$&$0.868\pm0.057$&$9.804\pm0.045$&$3.133\pm0.019$&$<0.033$&$-0.001\pm0.058$&$0$\\
$\Delta\alpha=-2^\circ$&$2.301\pm0.017$&$1.079\pm0.015$&$7.521\pm0.076$&$0.870\pm0.056$&$9.805\pm0.043$&$3.133\pm0.019$&$<0.035$&$-2.001\pm0.061$&$0$\\
$\Delta\alpha=-5^\circ$&$2.301\pm0.017$&$1.079\pm0.015$&$7.520\pm0.078$&$0.867\pm0.057$&$9.805\pm0.045$&$3.133\pm0.019$&$<0.037$&$-5.000\pm0.062$&$0$\\

\hline

\multicolumn{9}{c}{PLANCK, $r=0$, $\Delta\alpha\equiv0$}\\
$\Delta\alpha=0^\circ$&$2.300\pm0.017$&$1.079\pm0.015$&$7.518\pm0.078$&$0.870\pm0.057$&$9.806\pm0.044$&$3.133\pm0.019$&$<0.032$&$-$&$0$\\
$\Delta\alpha=-2^\circ$&$2.304\pm0.017$&$1.082\pm0.015$&$7.506\pm0.080$&$0.868\pm0.056$&$9.804\pm0.045$&$3.134\pm0.019$&$<0.048$&$-$&$6.98$\\
$\Delta\alpha=-5^\circ$&$2.327\pm0.017$&$1.085\pm0.015$&$7.499\pm0.077$&$0.858\pm0.056$&$9.836\pm0.044$&$3.119\pm0.019$&$0.081\pm0.058$&$-$&$261.68$\\

\hline

\multicolumn{9}{c}{PLANCK, $r=0.3$, $\Delta\alpha$ free}\\
$\Delta\alpha=0^\circ$&$2.300\pm0.017$&$1.080\pm0.015$&$7.512\pm0.078$&$0.870\pm0.056$&$9.801\pm0.045$&$3.135\pm0.020$&$0.305\pm0.100$&$0.003\pm0.060$&$0$\\
$\Delta\alpha=-2^\circ$&$2.300\pm0.016$&$1.080\pm0.015$&$7.513\pm0.077$&$0.871\pm0.056$&$9.802\pm0.044$&$3.135\pm0.020$&$0.308\pm0.100$&$-2.003\pm0.059$&$0$\\
$\Delta\alpha=-5^\circ$&$2.300\pm0.017$&$1.080\pm0.015$&$7.513\pm0.076$&$0.873\pm0.057$&$9.800\pm0.044$&$3.136\pm0.020$&$0.303\pm0.102$&$-4.998\pm0.059$&$0$\\

\hline

\multicolumn{9}{c}{PLANCK, $r=0.3$, $\Delta\alpha\equiv0$}\\
$\Delta\alpha=0^\circ$&$2.300\pm0.017$&$1.080\pm0.015$&$7.515\pm0.078$&$0.872\pm0.057$&$9.800\pm0.045$&$3.135\pm0.020$&$0.309\pm0.099$&$-$&$0$\\
$\Delta\alpha=-2^\circ$&$2.303\pm0.017$&$1.082\pm0.015$&$7.503\pm0.077$&$0.863\pm0.056$&$9.802\pm0.045$&$3.134\pm0.020$&$0.317\pm0.104$&$-$&$6.54$\\
$\Delta\alpha=-5^\circ$&$2.333\pm0.017$&$1.087\pm0.015$&$7.491\pm0.075$&$0.857\pm0.056$&$9.831\pm0.045$&$3.121\pm0.019$&$0.389\pm0.104$&$-$&$260.74$\\

\hline\hline
\end{tabular}}
\end{table}

Therefore, we simulate the future CMB power spectra with Planck to
investigate the effect of non-zero rotation angle on the constraints
of cosmological parameters. The fiducial model we choose is the
best-fit WMAP7 model \cite{WMAP7}: $\Omega_{b}h^2=0.023$,
$\Omega_{c}h^2=0.108$, $\Omega_\Lambda=0.752$, $\tau=0.087$,
$n_{s}=0.98$, and $\log{[10^{10}A_{s}]}=3.135$ at $k_{\rm
s0}=0.002\,$Mpc$^{-1}$. For the tensor to scalar ratio $r$, we have
two fiducial models: $r=0$ and $r=0.3$. For each model, we consider
three fiducial values of the rotation angle with
$\Delta\alpha=0,\,-2,\,-5$ deg. Here, we neglect the systematic
error of future CMB measurement and the CMB lensing effect. In Table
\ref{table:eCSfut} we list the constraints on the cosmological
parameters with different fiducial models.

\begin{figure}[htbp]
\begin{center}
\includegraphics[scale=0.31]{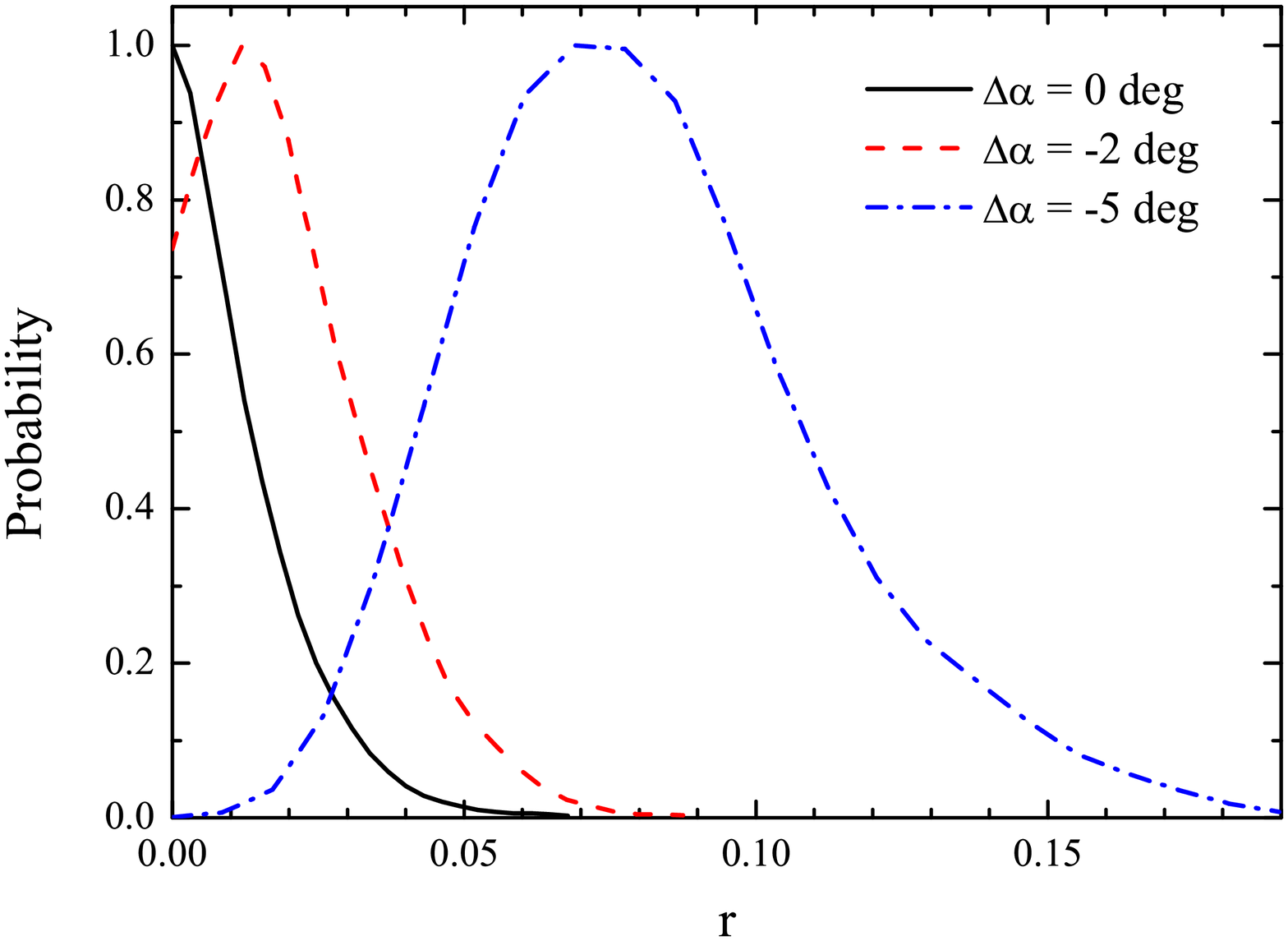}
\includegraphics[scale=0.31]{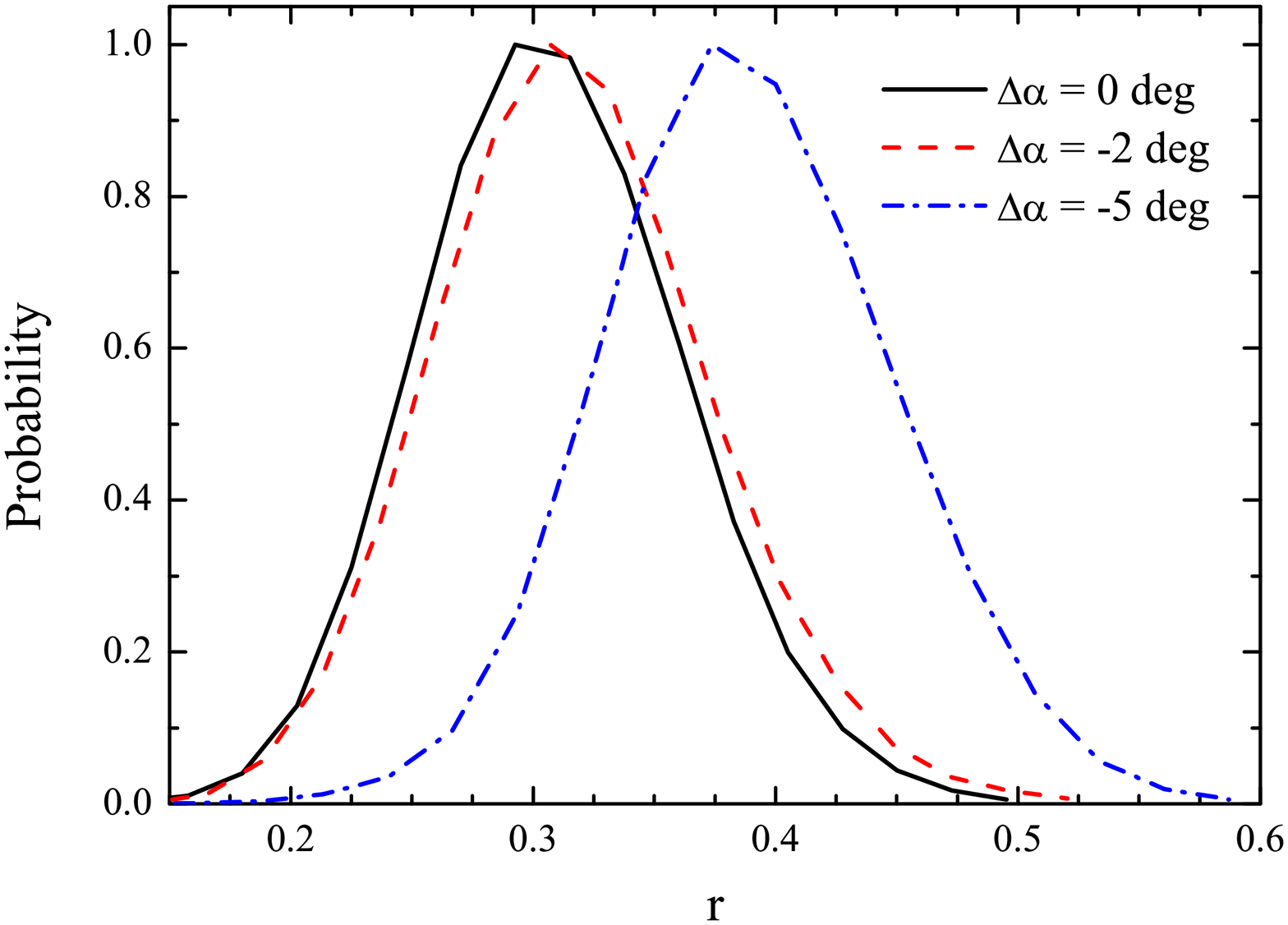}
\caption{The one-dimensional posterior distributions of the tensor
to scalar ratio $r$ derived from the future Planck mock data without
including the rotation angle. The fiducial values of $r$ are 0 (left
panel) and 0.3 (right panel). The rotation angles considered in the
fiducial model are: $\Delta\alpha=0$ deg (black solid lines),
$\Delta\alpha=-2$ deg (red dashed lines), and $\Delta\alpha=-5$ deg
(blue dash-dotted lines). \label{fig:tensorfut}}
\end{center}
\end{figure}

When we include the rotation angle in the calculations, as you can
see in Table \ref{table:eCSfut}, the fiducial values of input
parameters are always recovered. The future Planck measurement could
shrink the standard deviation of rotation angle by a factor of 15,
namely $\sigma(\Delta\alpha)\simeq0.06$ deg, which is consistent
with our previous results \cite{Xia:2008a,Xia:Planck}. The non-zero
rotation angle and the possible cosmological CPT violation can be
verified by the future CMB measurements.

However, if we neglect the effect of rotation angle at all
($\Delta\alpha\equiv0$), the contribution of non-zero rotation angle
on the BB power spectrum will be wrongly considered as the
primordial tensor perturbations. For the fiducial model $r=0$, the
BB power spectrum due to the primordial tensor B-mode should vanish.
But in the presence of non-zero fiducial rotation angle
($\Delta\alpha=-2,\,-5$ deg), the BB power spectrum should be
non-vanishing. If we force the rotation angle to be zero in the
analysis, the value of $r$ will be enlarged to match the mock
non-zero BB power spectrum. Using the $\Delta\alpha$-free cases as
the reference model, in Table \ref{table:eCSfut} we can see that the
minimal $\chi^2$ of the cases with $\Delta\alpha\equiv0$ are
extremely larger, which implies that this kind of model does not fit
the mock data well. In the left panel of Fig.\ref{fig:tensorfut}, we
find that the larger the fiducial value of $\Delta\alpha$ is, the
larger the obtained central value of $r$ becomes. Consequently, this
large value of $r$ will enhance the CMB TT power spectrum at large
scales. The contribution of primordial scalar perturbations on the
TT power spectrum will be suppressed, which will change constraints
of other cosmological parameters, for example, the baryon energy
density $\Omega_bh^2$ becomes larger and the amplitude of primordial
scalar perturbations $A_s$ becomes smaller. See the Table
\ref{table:eCSfut} for details. Therefore, if we do not take the
effect of rotation angle into account properly, the constraints of
cosmological parameters derived from the future CMB data with high
accuracy will be biased. We also find the similar situation for the
case with $r=0.3$ (right panel of Fig.\ref{fig:tensorfut}).

\subsection{Gravitational Chern-Simons Term}

In this subsection, we study the effect of gCS term on the CMB power
spectra and constrain the parameter $\epsilon$ from the CMB
polarization data. Note that, in our analyses we do not follow
Ref.\cite{Li:2009} which took the effects of eCS and gCS terms into
account simultaneously, since the current constraint on $\epsilon$
is still too weak to consider this kind of degeneracy. We always
assume the rotation angle $\Delta\alpha=0$ and neglect the
systematic effects of CMB polarization data in the calculations.
Therefore, we only have one more free parameter $\epsilon\in[-1,1]$,
induced by the gCS term, and assume that it is scale-independent.

\begin{figure}[t]
\begin{center}
\includegraphics[scale=0.31]{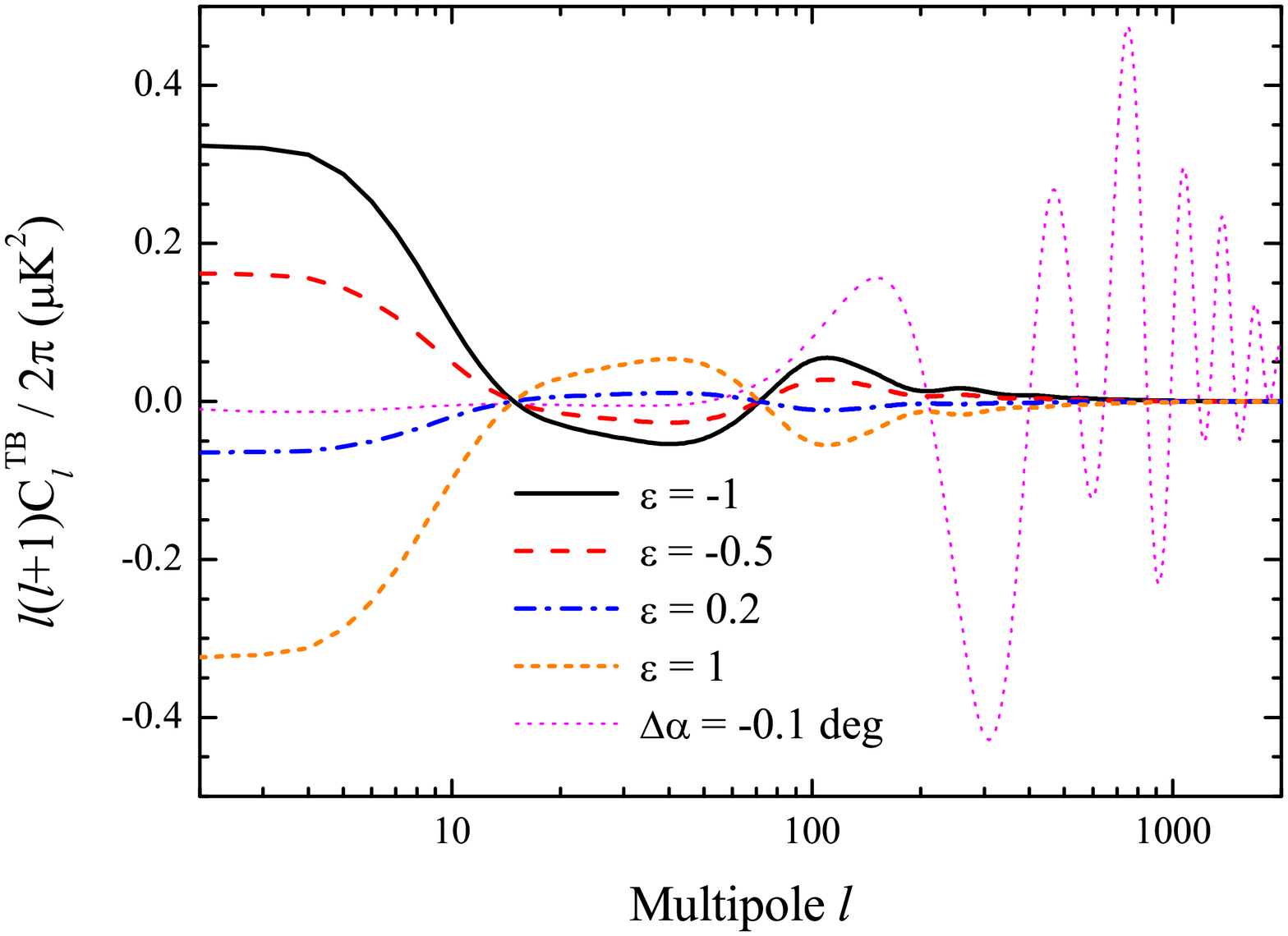}
\includegraphics[scale=0.31]{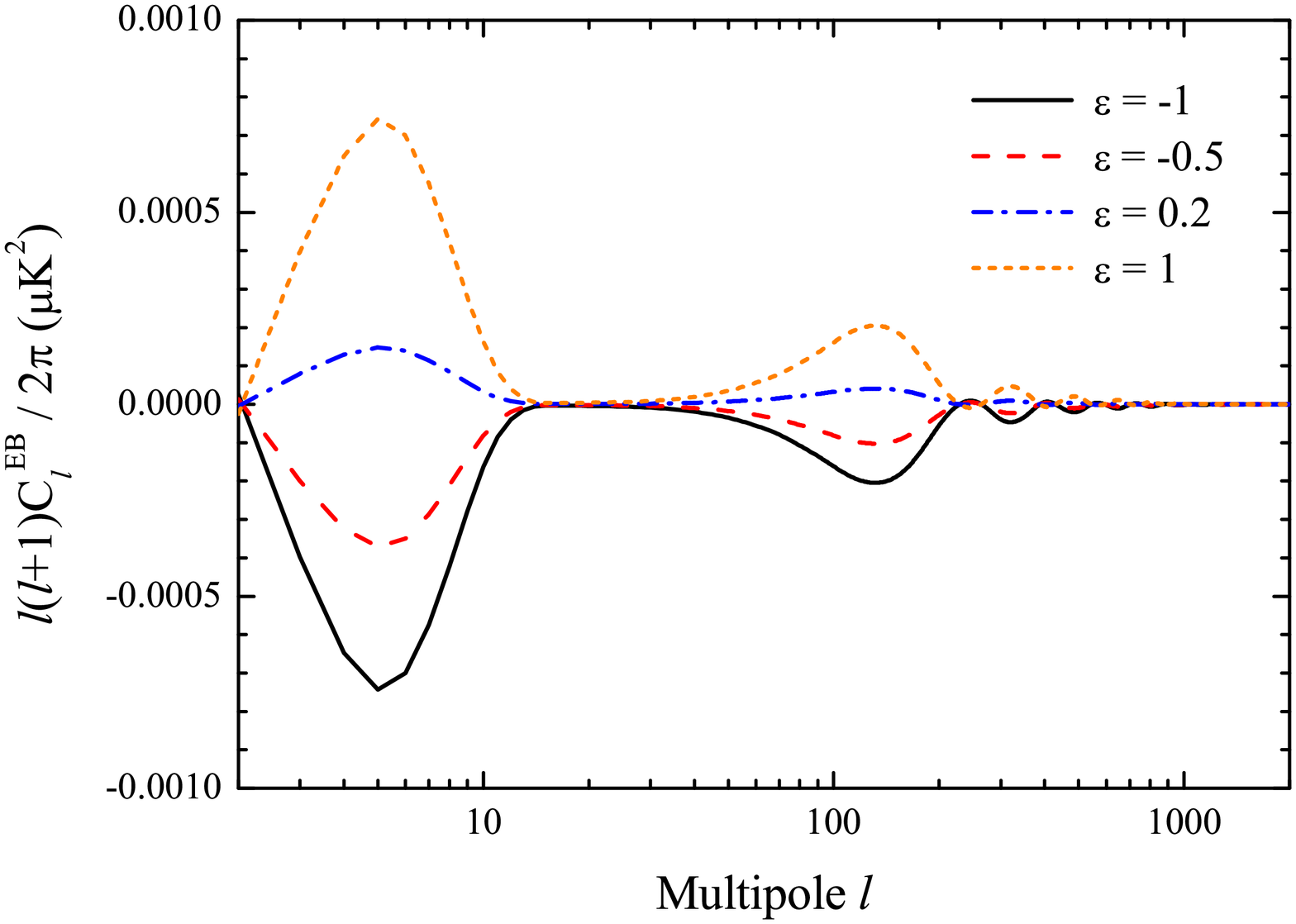}
\includegraphics[scale=0.31]{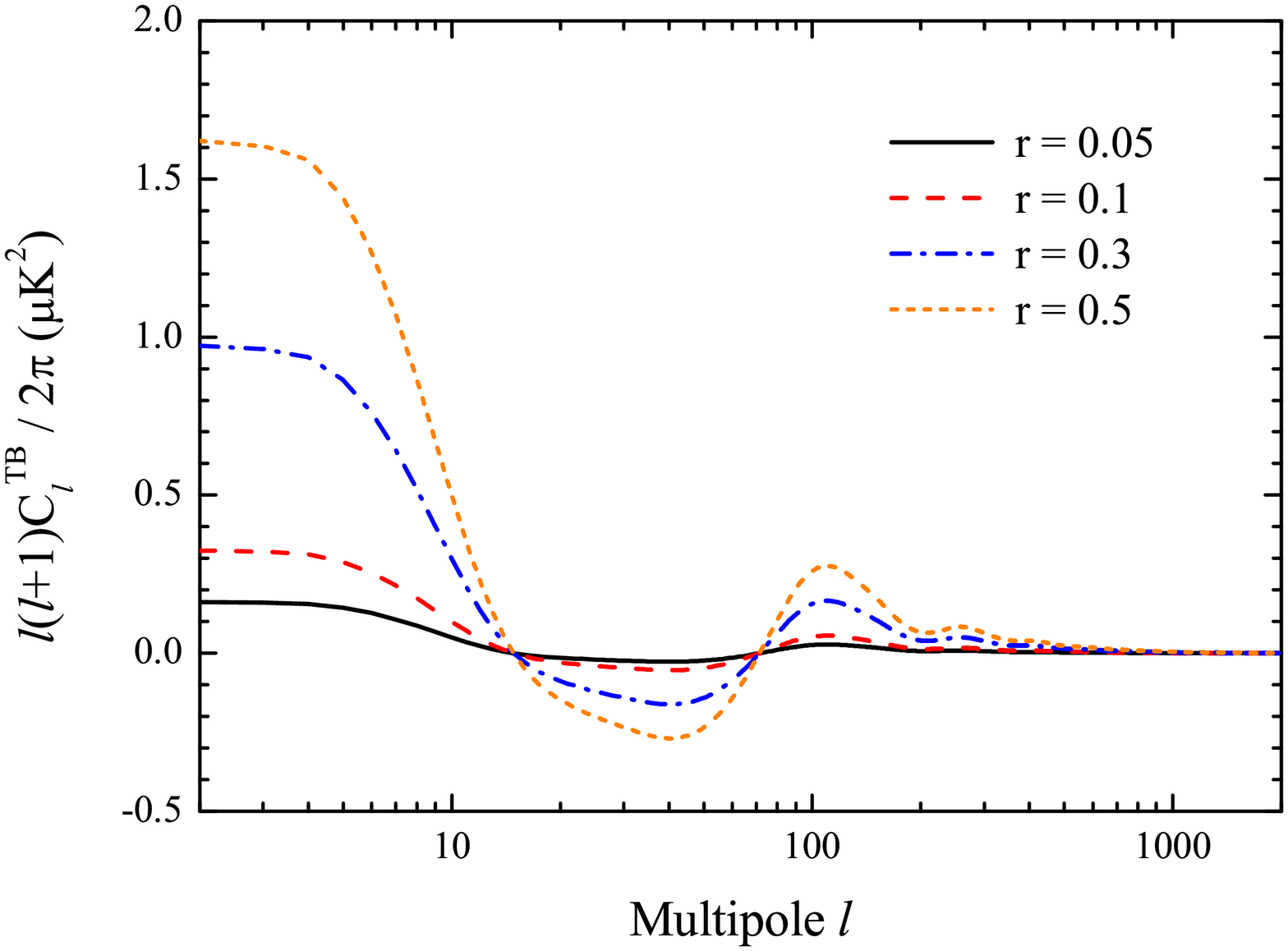}
\includegraphics[scale=0.31]{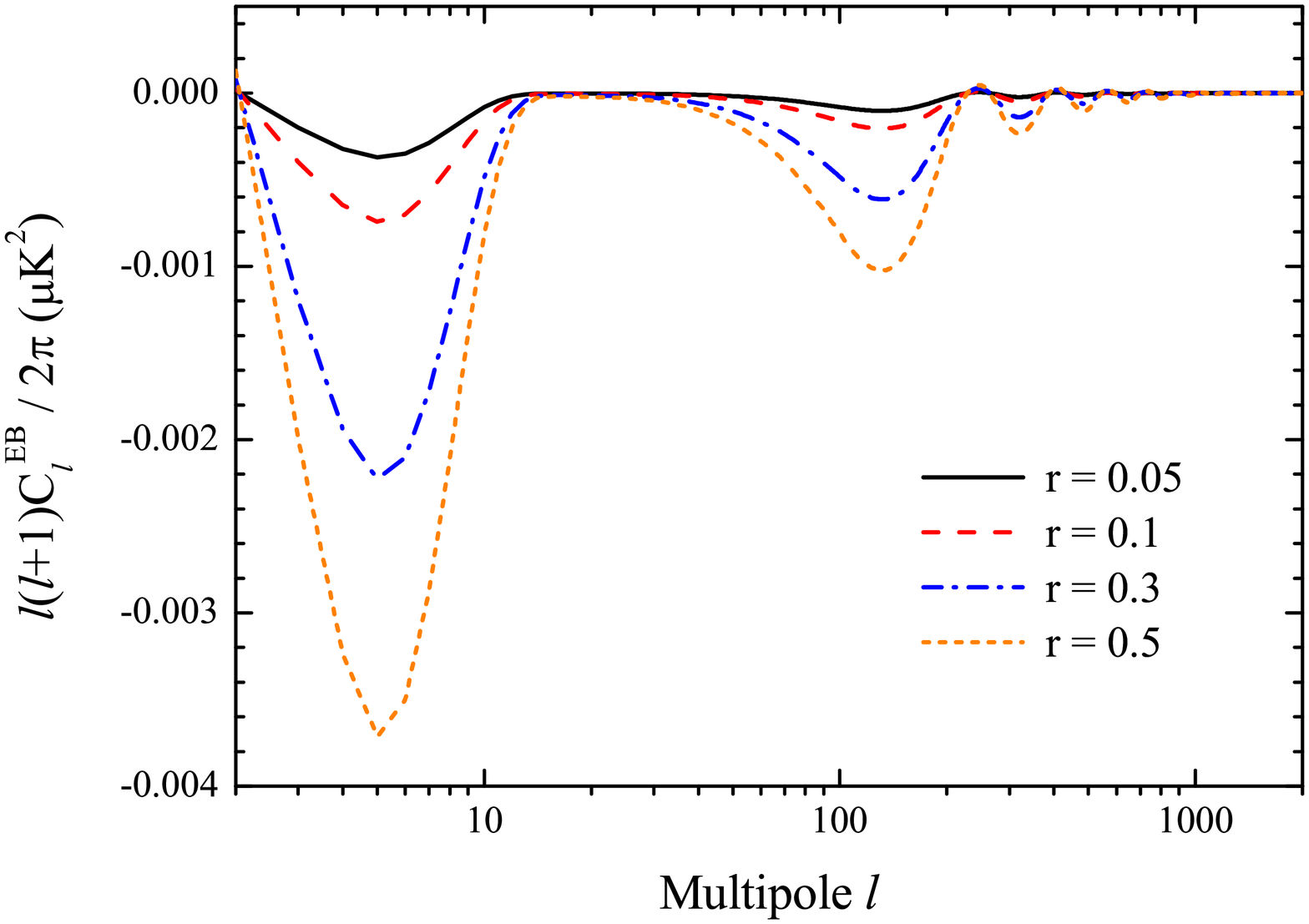}
\caption{Upper panels: The theoretical TB and EB power spectra for
different values of $\epsilon$ with $r=0.1$. Below panels: The
theoretical TB and EB power spectra for different values of $r$ with
$\epsilon=-1$. We also show the theoretical TB power spectrum with
the rotation angle $\Delta\alpha=-0.1$ deg for comparison in the
upper left panel (magenta dotted line).\label{fig:gCS}}
\end{center}
\end{figure}

Firstly, we show the effect of the gCS term on the CMB polarization
power spectra. As we discuss before, a non-zero $\epsilon$ only
generates the non-zero TB and EB power spectra and leaves other four
power spectra unchanged. In the upper panels of Fig.\ref{fig:gCS},
we plot the CMB TB and EB power spectra for different fiducial
values of $\epsilon$ with $r=0.1$. We find that the non-zero
$\epsilon$ mainly affect the TB power spectrum on large scales
($\ell < 200$). There is a significantly enhancement on the
amplitude of TB power spectrum on the largest scales ($\ell<10$),
which is related to the CMB reionization information
\cite{Saito:2007}. We also notice that there are only two crossing
points around $\ell \sim 15$ and 70 on the TB power spectrum. For
comparison, we show the TB power spectrum generated by a rotation
angle $\Delta\alpha=-0.1$ deg in the plot (magenta dotted line). As
we know, the non-zero rotation angle will generate the TB power
spectrum, based on the relation: $C_{\ell}^{\rm 'TB}=C_{\ell}^{\rm
TE}\sin(2\Delta\alpha)$ which implies that the shape of TB power
spectrum is exactly the same with that of TE power spectrum.
Therefore, the TB power spectrum with a non-zero $\Delta\alpha$ has
many crossing points at high multipoles ($\ell > 200$) and its
signal mainly comes from the small scales. We could distinguish the
TB power spectrum from the eCS term and that from the gCS term
clearly on small scales. Moreover, in the plot we only show the TB
power spectrum with $\Delta\alpha=-0.1$ deg which is much smaller
than the best fit value from WMAP7+B03+BICEP data,
$\Delta\alpha\sim-2$ deg, because the signal of TB power spectrum
from the eCS term is much higher than that from gCS term, especially
on small scales. Thus, if we find very large signal on small scales
of TB power spectrum from the future observation, it only could be
the signature of a non-zero rotation angle.

We find the similar situation on the EB power spectrum. From the
upper right panel of Fig.\ref{fig:gCS}, we can see that the effect
of the EB power spectrum, induced by the gCS term, is mainly on
large scales and becomes to zero at high multipoles. There are many
crossing points on the EB power spectrum, while the EB power
spectrum, induced by the eCS term, does not have the crossing point.
Based on the relation: $C_{\ell}^{\rm 'EB} \sim (C_{\ell}^{\rm
EE}-C_{\ell}^{\rm BB})\sin(4\Delta\alpha)$, the EB power spectrum
can not cross the zero line, since the EE power spectrum is always
positive and larger than the BB power spectrum. Here, we do not show
the curve in the plot for comparison, since the amplitude of EB
power spectrum produced by the eCS term is much larger than that by
the gCS term. Thus, the EB power spectrum can also be used to
distinguish the eCS and gCS models, if the future CMB polarization
data are accurate enough.

\begin{figure}[t]
\begin{center}
\includegraphics[scale=0.5]{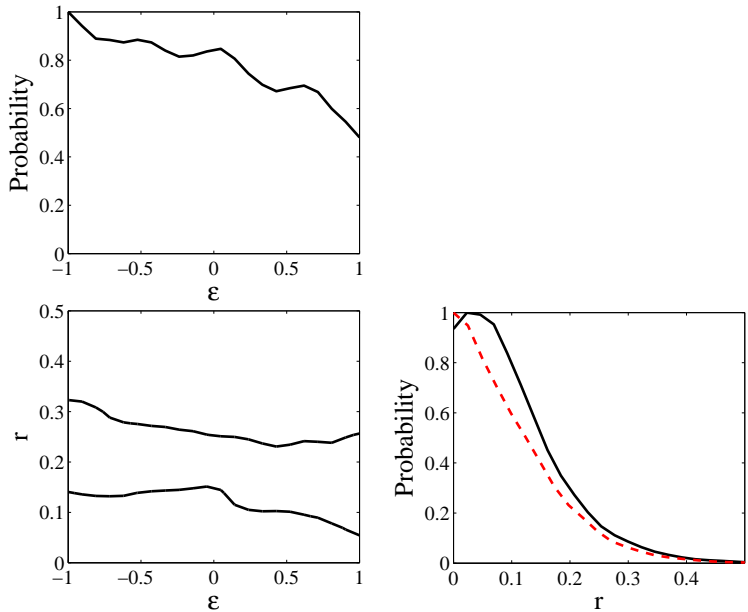}
\caption{Marginalized one-dimensional and two-dimensional
distributions ($1,\,2\,\sigma$ contours) of $\epsilon$ and $r$. In
the right below panel we also show the constraint of $r$ without
considering the non-zero gCS term, $\epsilon=0$ (red dashed line).
\label{fig:gCScur}}
\end{center}
\end{figure}

In the below two panels of Fig.\ref{fig:gCS}, we show the dependence
of $r$ on the TB and EB power spectra with $\epsilon=-1$. One can
see that the amplitudes of TB/EB power spectra are highly related to
by the fiducial value of $r$, and their shapes remain unchanged.
Therefore, the significance of detection of non-zero $\epsilon$ will
be directly determined by the primordial tensor perturbations $r$.
At present, CMB observations do not detect the signature of non-zero
$r$, which means that results on $r$ are consistent with zero.
Consequently, we can not constrain the parameter $\epsilon$ very
well from the current CMB polarization data, due to the very small
primordial tensor perturbations. In Fig.\ref{fig:gCScur}, we show
the constraints on $\epsilon$ and $r$ from the WMAP7+B03+BICEP+QUaD
data combination after marginalized over other cosmological
parameters. From the one-dimensional distribution of $\epsilon$, as
well as the flat two-dimensional contours, apparently the present
CMB data can not give good constraints on $\epsilon$. There is a
peak around $\epsilon=-1$, which implies that a right-hand polarized
gravitational wave might be slightly favored. We also show the
one-dimensional distribution of $r$ including and not including the
free parameter $\epsilon$. Due to the degeneracy between $\epsilon$
and $r$, the constraint on $r$ is slightly relaxed when a free
$\epsilon$ is included in the calculation.

\begin{figure}[t]
\begin{center}
\includegraphics[scale=0.6]{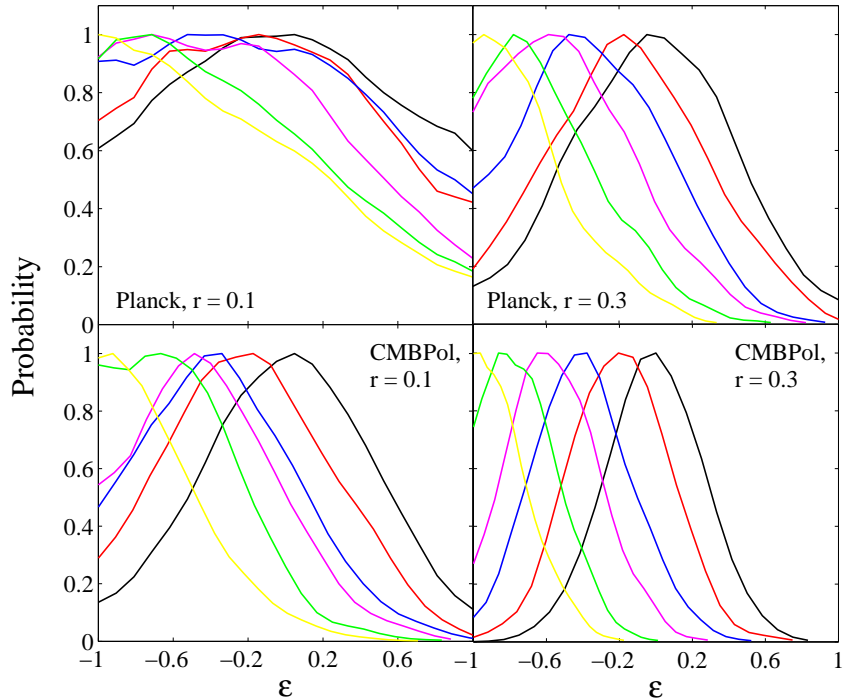}
\caption{The one-dimensional posterior distributions of $\epsilon$
derived from the future Planck (upper panels) and CMBPol (below
panels) measurements with different fiducial models of $r$. For each
case, we have six fiducial values of $\epsilon$: $\epsilon=0$
(black), $-0.2$ (red), $-0.4$ (blue), $-0.6$ (magenta), $-0.8$
(green), and $-1$ (yellow). \label{fig:gCSfut}}
\end{center}
\end{figure}

Thus, we use the same fiducial model in the last subsection to
simulate the future Planck polarization data to constrain
$\epsilon$. For the tensor to scalar ratio $r$, we have two fiducial
models: $r=0.1$ and $r=0.3$. For each model, we consider six
fiducial values of $\epsilon$ with
$\epsilon=0,\,-0.2,\,-0.4,\,-0.6,\,-0.8,\,-1$. In Table
\ref{table:gCSfut} and upper panels of Fig.\ref{fig:gCSfut} we show
the constraints on $\epsilon$ from the future CMB mock data with
different fiducial models. If we choose the fiducial value $r=0.1$,
even the Planck mock data can not give good constraint on $\epsilon$
(the upper left panel). The 95\% confidence levels of $\epsilon$ are
always $[-1,1]$ for six fiducial values of $\epsilon$. As we discuss
before, the larger the value of $r$ is, the higher the amplitudes of
CMB TB/EB power spectra are (the below panels of Fig.\ref{fig:gCS}).
When the fiducial value of $r$ becomes larger ($r=0.3$), the
constraints on $\epsilon$ become better. The standard deviations of
$\epsilon$ have been shrunk significantly. But the constraints are
still not conclusive. Obviously, the future Planck measurement will
be very difficult to constrain $\epsilon$, if the primordial tensor
perturbations are small, $r<0.1$. Therefore, we also simulate the
future CMBPol polarization data with higher accuracy and re-do the
calculations, see Table \ref{table:eCSfut} and below panels of
Fig.\ref{fig:gCSfut}. In this case, even with $r=0.1$ the CMBPol
data could also give good constraints on $\epsilon$. Finally, we
obtain the standard deviation of $\epsilon$:
$\sigma(\epsilon)\sim0.1$ with the fiducial value $r=0.3$. The
future CMBPol experiment could constrain $\epsilon$ better.

\begin{table}[t]
\caption{Constraints on $\epsilon$ (95\% C.L.) from future CMB mock
data.}\label{table:gCSfut}
\begin{tabular}{c|c|c|c|c}
\hline \hline

Fiducial $\epsilon$ & Planck, $r=0.1$ & Planck, $r=0.3$ & CMBPol, $r=0.1$ & CMBPol, $r=0.3$\\

\hline

$\epsilon=0$   &$[-1,1]$&$[-0.69,0.66]$&$[-0.71,0.70]$&$[-0.43,0.42]$\\
$\epsilon=-0.2$&$[-1,1]$&$[-1,0.57]$   &$[-1,0.54]$  &$[-0.61,0.23]$\\
$\epsilon=-0.4$&$[-1,1]$&$[-1,0.32]$  &$[-1,0.37]$  &$[-0.80,0.05]$\\
$\epsilon=-0.6$&$[-1,1]$&$[-1,0.22]$  &$[-1,0.25]$  &$[-1,-0.12]$\\
$\epsilon=-0.8$&$[-1,1]$&$[-1,0.02]$  &$[-1,0.03]$  &$[-1,-0.36]$\\
$\epsilon=-1.0$&$[-1,1]$&$[-1,-0.12]$  &$[-1,-0.08]$  &$[-1,-0.48]$\\

\hline  \hline
\end{tabular}
\end{table}


\section{Conclusions and Discussions}\label{summary}

Probing the violation of fundamental symmetries is an important way
to search for the new physics beyond the standard model. In this
paper we present constraints on the eCS and gCS models using the
latest CMB polarization data, as well as the future simulated mock
data.

In the constraints of rotation angle, induced by the eCS model, we
extend our previous works by including the systematic errors of CMB
polarization measurements. We consider two rotation angles,
$\Delta\alpha$ and $\beta$, in order to take into account the real
rotation signal and a systematic error for each CMB polarization
measurement, and impose priors on the systematic errors. Adding the
systematic effects of CMB polarization data, we do not find
significant change on the constraints of $\Delta\alpha$, except the
error bars become slightly larger. WMAP7+B03+BICEP data combination
still favors a non-zero rotation angle about $2.2\,\sigma$
confidence level, namely $\Delta\alpha=-2.28\pm1.02~({\rm deg})$, no
matter whether the systematic effects are included. We still find a
$\sim2\,\sigma$ tension between QUaD and WMAP7+B03+BICEP
observations. When combining all CMB polarization data together, we
obtain the tight constraint on the rotation angle at 95\% confidence
level: $-1.34<\Delta\alpha<0.82$ (deg), which is consistent with a
CPT-conserving Universe.

Since the current constraints on the rotation angle are not
conclusive, we simulate the future Planck polarization data. We find
that the future CMB data could significantly improve the constraint
of $\Delta\alpha$ by a factor of $15$
($\sigma(\Delta\alpha)\simeq0.06$ deg). Because a non-zero rotation
angle will generate the BB power spectrum, we need to take this
effect into account properly in the future data analysis. Otherwise
the constraints of cosmological parameters will be apparently
biased.

We also give the constraint on the parameter $\epsilon$, which is
induced by the gCS term and denotes the difference between the
right- and left-handed polarized components. Since the effects of
non-zero $\epsilon$ on the TB and EB power spectra are very small,
the parameter $\epsilon$ is almost unconstrained from the current
polarization data. Therefore, we simulate the future Planck and
CMBPol data to constrain $\epsilon$ further. We find that the future
Planck data is very difficult to constrain $\epsilon$, if the
primordial tensor perturbations are small, $r<0.1$. Using the future
CMBPol data, the constraint of $\epsilon$ can be improved to
$\sigma(\epsilon)\sim 0.1$, if the fiducial value $r=0.3$. We could
use the future CMBPol data to verify the gCS model.


\section*{Acknowledgements}

We acknowledge the use of the Legacy Archive for Microwave
Background Data Analysis (LAMBDA). Support for LAMBDA is provided by
the NASA Office of Space Science.


\end{document}